\documentclass[twocolumn]{aastex63}
\usepackage{graphicx}
\usepackage{blindtext}
\usepackage{amsmath}
\usepackage{mathtools}
\usepackage{multirow}
\usepackage{hyperref}
\hypersetup{
	colorlinks	= true,
	linkcolor	= red,
	urlcolor	= cyan,
	citecolor	= blue
}
\usepackage{longtable}
\usepackage{makecell}
\usepackage{lineno}
\usepackage{array}
\newcolumntype{C}[1]{>{\centering\arraybackslash}p{#1}}
\newcolumntype{X}[1]{>{\centering\let\newline\\\arraybackslash}p{#1}}

\newcommand{\exoclock}{\mbox{ExoClock}}

\newcommand{\uselinenumbers}{}
\newcommand{\markchange}[1]{#1}

\newcommand{\tabparameterssfile}{tab_parameters.tex}

\begin{document}

\uselinenumbers

\title{ExoClock Project IV: A homogeneous catalogue of 620 updated exoplanet ephemerides}

\correspondingauthor{A. Kokori}

\email{anastasia.kokori.19@ucl.ac.uk}

\author{A. Kokori}
\affiliation{University College London, Gower Street, London, WC1E 6BT, UK}

\author{A. Tsiaras}
\affiliation{Department of Physics, Aristotle University of Thessaloniki, University Campus, Thessaloniki, 54124, Greece}

\author{G. Pantelidou}
\affiliation{Department of Physics, Aristotle University of Thessaloniki, University Campus, Thessaloniki, 54124, Greece}

\author{A. Jones}
\affiliation{Amateur Astronomer\footnote{A list of associated private observatories that contributed to this work can be found in Appendix A}}\affiliation{British Astronomical Association, PO Box 702, Tonbridge TN9 9TX, UK}

\author{A. Siakas}
\affiliation{Department of Physics, Aristotle University of Thessaloniki, University Campus, Thessaloniki, 54124, Greece}

\author{B. Edwards}
\affiliation{AIM, CEA, CNRS, Université Paris-Saclay, Université de Paris, F-91191 Gif-sur-Yvette, France}\affiliation{University College London, Gower Street, London, WC1E 6BT, UK}

\author{G. Tinetti}
\affiliation{University College London, Gower Street, London, WC1E 6BT, UK}

\author{A. Wünsche}
\affiliation{Observatoire des Baronnies Provençales, Route de Nyons, 05150 Moydans, France}

\author{Y. Jongen}
\affiliation{Amateur Astronomer\footnote{A list of associated private observatories that contributed to this work can be found in Appendix A}}\affiliation{Observatoire de Vaison-La-Romaine, Départementale 51, près du Centre Equestre au Palis - 84110 Vaison-La-Romaine, France}

\author{F. Libotte}
\affiliation{Amateur Astronomer\footnote{A list of associated private observatories that contributed to this work can be found in Appendix A}}\affiliation{Agrupació Astronòmica de Sabadell, Carrer Prat de la Riba, 116, 08206 Sabadell, Barcelona, Spain}\affiliation{Groupe Européen d'Observations Stellaires (GEOS), Bailleau l'Evéque, France}\affiliation{Instituto de Astrofísica de Canarias (IAC), E-38200 La Laguna,  Tenerife, Spain}

% ...

\begin{abstract}
The ExoClock project is an open platform aiming to monitor exoplanets by integrating observations from space and ground based telescopes. This study presents an updated catalogue of 620 exoplanet ephemerides, integrating 30000 measurements from ground-based telescopes (the ExoClock network), literature, and space telescopes (Kepler, K2 and TESS). The updated catalogue includes 277 planets from TESS which require special observing strategies due to their shallow transits or bright host stars. This study demonstrates that data from larger telescopes and the employment of new methodologies such as synchronous observations with small telescopes, are capable of monitoring special cases of planets. The new ephemerides show that 45\% of the planets required an update while the results show an improvement of one order of magnitude in prediction uncertainty. The collective analysis also enabled the identification of new planets showing TTVs, highlighting the importance of extensive observing coverage. Developed in the context of the ESA’s Ariel space mission, with the goal of delivering a catalogue with reliable ephemerides to increase the mission efficiency, ExoClock’s scope and service have grown well beyond the remit of Ariel. The ExoClock project has been operating in the framework of open science, and all tools and products are accessible to everyone within academia and beyond, to support efficient scheduling of future exoplanet observations, especially from larger telescopes where the pressure for time allocation efficiency is higher (Ariel, JWST, VLT, ELT, Subaru etc). The inclusion of diverse audiences in the process and the collaborative mode not only foster democratisation of science but also enhance the quality of the results. 
\end{abstract}


\section{INTRODUCTION}

Until today, more than 5900 exoplanets have been detected, and although discoveries of new exoplanets continue daily thanks to facilities such as TESS \citep{Ricker2014}, we have now entered the characterization era of exoplanets. A dedicated characterization survey will be conducted by the Ariel mission \citep{Tinetti2018}, which aims to study 1000 exoplanet atmospheres. Ariel will observe thousands of transits and eclipses of exoplanets to further investigate their composition and their nature. 

The properties of the planets and their host stars need to be precisely determined before Ariel is launched to increase the efficiency of the mission. For example, well-determined stellar ages are important in constraining planetary properties and revealing the composition of giant planets \citep{Muller2023}. Other important parameters are stellar masses, radii, temperatures, elemental abundances, and activity indices, for which \cite{Danielski2022} and \cite{Magrini2022} aim to develop a homogeneous catalogue. Similarly, important planetary parameters include the planetary mass, radius, temperature, transit duration, transit depth, and transit timing -- i.e. the ephemeris.

Precision in transit timings is crucial to increase the mission efficiency and, therefore, avoid wasting significant observing time of Ariel. Predictions can be inaccurate for several reasons, including ephemeris uncertainties \citep{2019AnA...622A..81M}, and bias in the initial period or mid-time. Inaccuracies can also arise due to Transit Timing Variations (TTVs) that occur as a result of physical processes such as orbital decay, orbital precession, and planet-planet interactions \citep{Agol2018}.
 
At the same time, several ground-based telescopes are occupied with following up TESS candidates to confirm them as planets \citep{Collins2018,2023ApJS..265....1Y,Schulte2024} without further long-term monitoring. As a result, the ephemerides of these planets are not precise after their validation because, as highlighted in \cite{Narita2019}, the monitoring duration of TESS in one each sector is only 27 days, while, for comparison, the monitoring duration for Kepler was over 4 years.

\begin{figure}
\centering
\includegraphics[width=\columnwidth]{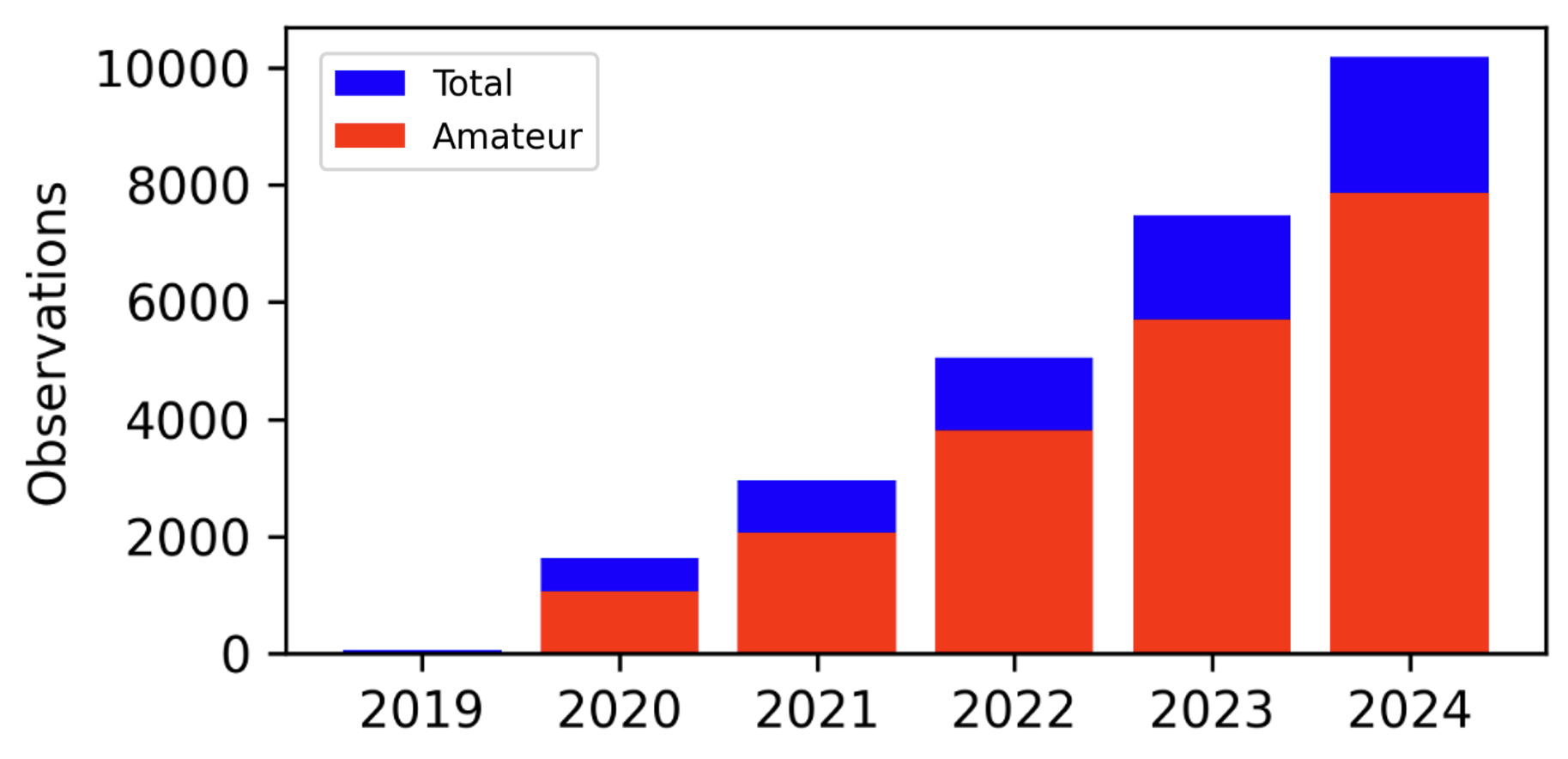}
\caption{Cumulative distribution of observations published by the \exoclock\ network.}
\label{fig:exoclock_observations}
\end{figure}

The effort of maximising the capabilities of Ariel observing plan started in 2019 with the launch of the \exoclock\ project \citep{exoclock1}, a dedicated, open, and integrated project with the aim of providing a consistent catalogue of ephemerides for all the Ariel candidate targets \citep{Edwards2022} by 2029. The \exoclock\ project has been operating for almost six years now, and so far it has been concluded that approximately 40\% of the initial ephemerides had to be updated due to significant uncertainties or biases \citep{exoclock2, exoclock3}. The number of observations has been increasing since the beginning of the project, surpassing 10000 in 2024 (Fig. \ref{fig:exoclock_observations}), while most of the observations are carried out with small- and medium-scale telescopes by amateur observers (73\%).

The \exoclock\ target list includes all the known Ariel candidate targets. This target list is constantly updated with newly confirmed planets, with the large majority of new planets coming from TESS. TESS has been operating since 2019 with more than 500 confirmed exoplanets and thousands of candidates that wait for confirmation \citep{Ricker2014,Collins2018,Magliano2023}, and the current \exoclock\ target list contains 277 planets discovered by TESS (36\% of the total current catalogue of 775 planets). The addition of the TESS discoveries has altered the characteristics of the target list in two ways: firstly, these targets have on average lower S/N compared to planets discovered by ground-based instruments, and secondly, these targets have more planetary companions. More specifically, 61\% of the new planets discovered by TESS require telescopes larger than 16 inches for their follow-up observations, while this drops to 21\% in the rest of the target list (based on the sensitivity study in \cite{exoclock3}). Furthermore, 42\% of the new planets discovered by TESS belong to multi-planetary systems, while this drops to 18\% in the rest of the target list.

With the constant addition of new planets by the TESS mission and the particularity of these planets, it becomes apparent that the \exoclock\ project needs to continue monitoring known exoplanets to decrease the uncertainties and the biases on their ephemerides. Moreover, there is now the need to expand the sample of accessible targets by developing new observing techniques, utilizing larger ground-based and space-based telescopes, and integrating more external archives, in order to provide a reliable catalogue when Ariel launches in 2029. 

While the \exoclock\ project was developed in the context of the Ariel space mission, its importance and applicability has extended beyond the mission. The products of the project are being used by several research teams already, demonstrating that a homogeneous catalogue with reliable exoplanet parameters is important for the entire academic community.

In this study, we have used diverse resources of data, both from ground- and space-based telescopes, as well as mid-time points derived from literature studies. In addition, we describe how we have used multiple small telescopes to observe low S/N transits. In total, approximately 30000 mid-time points have been used to improve the ephemerides of 620 planets by decreasing biases, increasing accuracy and also detecting long-term phenomena (TTVs).

The open and integrated nature of the \exoclock\ project enables the efficient monitoring of the Ariel candidate targets, but it is also a successful vehicle for effective public engagement, where members of society (amateur and professional astronomers, as well as university and school students and members of the general public) participate actively in the scientific processes and contribute to a future space mission. In addition, universities and schools employ the \exoclock\ project for additional research projects, demonstrating that the framework of the \exoclock\ project provides research and training potential beyond the main scope of monitoring the ephemerides of planets for Ariel.

\section{DATA}

In this work we used light-curves from diverse resources including: the \exoclock\ network, the ASTEP observatory, the MuSCAT2 camera on the Telescopio Carlos Sánchez \citep{Narita2019}, the Las Cumbres Observatory \citep[LCO,][]{lco} telescopes, the Exoplanet Transit Database \citep[ETD,][]{etd}, the STScI Mikulski Archive for Space Telescopes (MAST) for the Kepler \citep{kepler}, K2 \citep{k2}, and TESS \citep{Ricker2014} space missions, as well as mid-transit times from the literature, to update the ephemerides of 620 exoplanets. All light-curves were acquired before the end of 2023, and the literature mid-transit times were published by December 2023. 

We performed the analysis of all the light-curves using the stellar and planetary parameters included in the Exoplanet Characterization Catalogue (ECC, Appendix \ref{sec:sup-info}), a dedicated catalogue within the \exoclock\ project \citep{exoclock1}, and the open-source Python package PyLightcurve \citep{Tsiaras2016B2016ApJ...820...99T}. In summary, the steps applied for each light-curve included:
\begin{itemize}
    \item calculation the limb-darkening coefficients \citep[ExoTETHyS,][]{Morello2020, exotethys_joss} using the Phoenix stellar models \citep{exotethys_phoenix}
    \item conversion of any time format to Barycentric Julian Date (BJD$_\mathrm{TDB}$)
    \item preliminary fitting using the Nelder-Mead minimisation \citep[SciPy,][]{Virtanen2020} of a transit model multiplied by a trend model
    \item removal of 3$\sigma$ outliers 
    \item scaling of the provided uncertainties based on the RMS of the normalised residuals
    \item Markov Chain Monte Carlo (MCMC) optimisation \citep[{\tt{emcee}},][]{ForemanMackey2013} leaving as free parameters only the $R_p/R_s$, the transit mid-time and the de-trending parameters
\end{itemize}

The light-curves from the \exoclock\ network were de-trended using a linear function of airmass or time and for more difficult cases, a second-order polynomial with time. The light-curves from Kepler, K2 and TESS all de-trended using a second-order polynomial with time.

After the fitting, we performed a quality evaluation individually for each light-curve. Light-curves that did not fulfil one or more of the criteria below (for more details see \cite{exoclock1}) were excluded further analysis.

\begin{itemize}
    \item autocorrelation and shapiro statistic indicate gaussian residuals at a 3$\sigma$ level
    \item transit signal-to-noise ratio ($ Depth/\sigma_{Depth}$) is above three,
    \item $Rp/Rs$ differs less than 3$\sigma$ from the literature value (for the ExoClock and ETD observations), or the weighted average of the mission (for the space observations),
    \item O-C value is in 3$\sigma$ agreement with other observations at similar time ($\sim$a month).
\end{itemize}

The final list of 620 planets includes those planets for which we collected data points at three or more different epochs and for which we could determine an ephemeris of better or similar quality to the initial ephemeris. A summary of all light-curves can be found in Table \ref{tab:data}.

\begin{figure}
\centering
\includegraphics[width=\columnwidth]{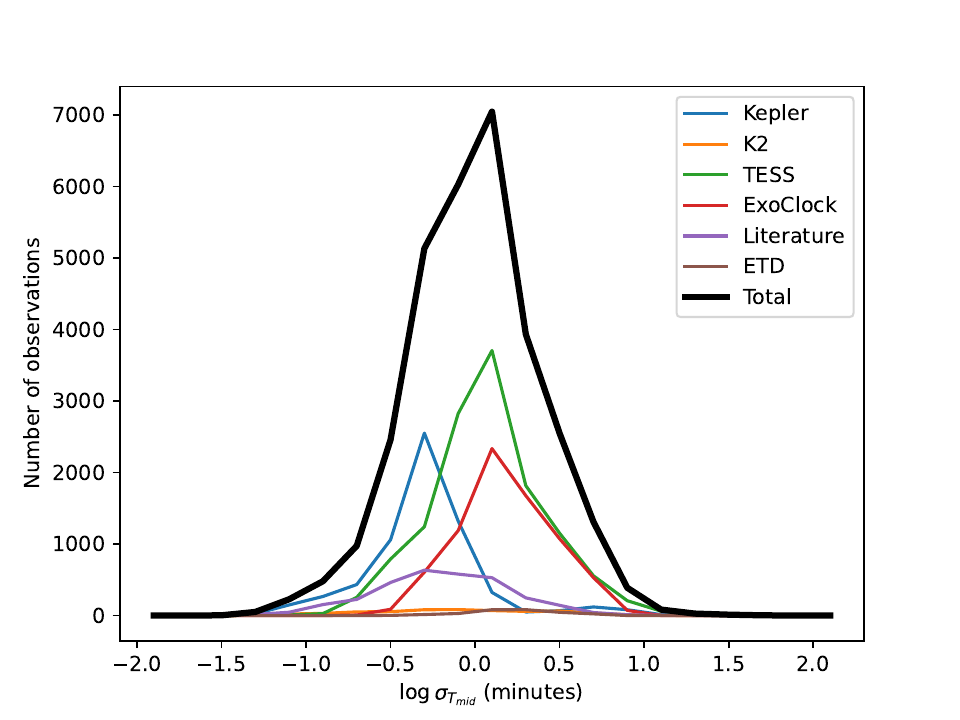}
\caption{Distribution of transit mid-time uncertainties among the different sources.}
\label{fig:precision}
\end{figure}

\begin{figure}
\centering
\includegraphics[width=\columnwidth]{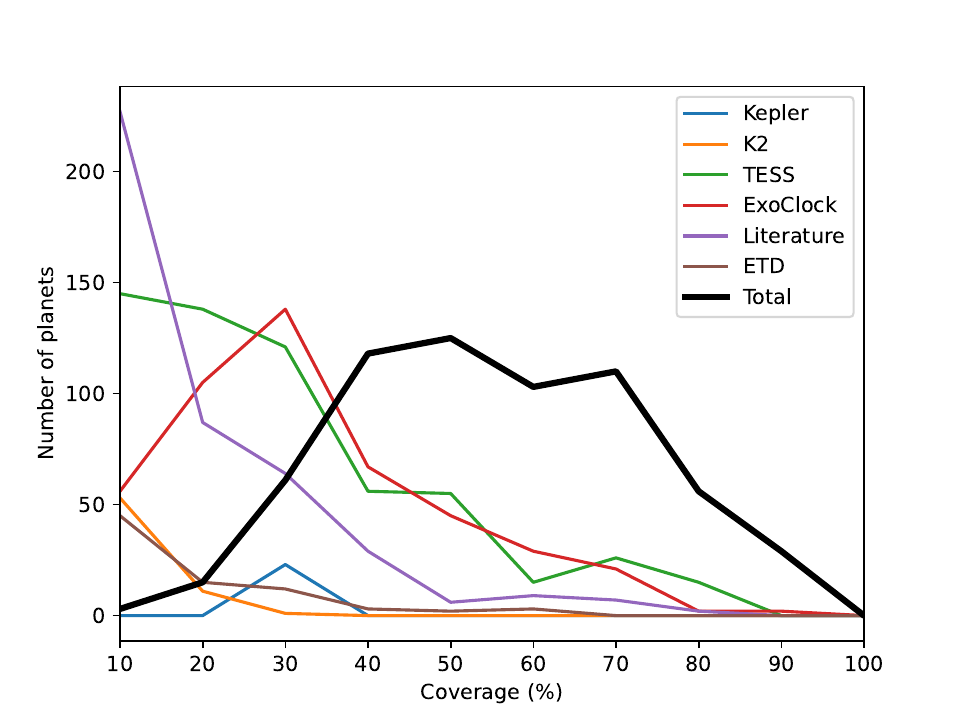}
\caption{Distribution of coverage among the different sources. As coverage we define the percentage of years (since the first observation in the database) for which at least one observation exists.}
\label{fig:coverage}
\end{figure}

Figures \ref{fig:precision} and \ref{fig:coverage} demonstrates the distributions of the precision and the coverage of the transit mid-times that were integrated to produce the final ephemerides. We define coverage as the the percentage of years since discovery for which at least one observation exists. We need to note here that 99\% of the observations used have transit mid-time uncertainties lower than 10 minutes, and that the median coverage of all sources combined together is 50\%, while individual sources do not reach more than 29\%.

\begin{table*}
\centering
\caption{Summary of the observations used in this work. As coverage we define the percentage of years (since the first observation in the database) for which at least one observation exists.}
\label{tab:data}
\begin{tabular}{c | c c c c c c c}
\hline
 							& ExoClock 	& ETD 		& Kepler 		& K2 		& TESS 		& Literature  		& \textbf{Total}			\\ \hline
Data points					& 7316		& 181		& 6471		& 572		& 12695		& 3109			& \textbf{30344}		\\ 
Years						& 2007-2023	& 2001-2021	& 2009-2013	& 2014-2018	& 2018-2023	& 2004-2023		& \textbf{2001-2023}		\\
Planets						& 466		& 40			& 23			& 65			& 573		& 431			& \textbf{620}			\\
Median $\sigma_{T_{mid}}$ 		& 1.4 min		& 1.7 min		& 0.5 min		& 0.7 min		& 1.2 min		& 0.6 min			& \textbf{1.0 min}		\\
Median coverage				& 27\%		& 14\%		& 24\%		& 10\%		& 22\%		& 13\%			& \textbf{50\%}			\\ \hline

\end{tabular}
\end{table*}

\subsection{Data from the ExoClock network}

In this work, we used 7588 light-curves from the \exoclock\ network which currently consists of 1700 participants -- 80\% of whom are amateur astronomers -- and 1600 telescopes. The telescope sizes range from 6 to 60 inches ($\sim$ 15 to 150 cm) and most of them ($\sim$80\%) are smaller than 17 inches, a number that highlights the power of small telescopes. \markchange{Figure \ref{fig:exoclock_observations} illustrates the distribution of the observations throughout the years since the start of the project.} 

The organisation of the \exoclock\ network is designed in a way to achieve maximum coverage of the planets and to ensure homogeneity in the results. The strategy behind the organisation of the project is described in detail in \citep{exoclock1}.

\subsection{Data from the MAST Archive}

Following the recipes in \cite{exoclock3}, in this work we also integrated light-curves from the Kepler \citep[long cadence,][]{kepler_data}, K2 \citep{k2_data} and TESS \citep[long cadence,][]{tess_data} missions,  acquired before the end of 2023. We included a time-span of one transit duration before and after each event, and we only considered those light-curves that were at least 80\% complete, both in-transit and out-of-transit -- i.e. total exposure time more than 0.8 times the transit duration before, during and after the transit. Table \ref{tab:parameters} includes the adjusted $a/R_s$ values, which are marked with an asterisk.

\subsection{Data from the MuSCAT2 camera on the Telescopio Carlos Sánchez}

The MuSCAT2 instrument, is a unique system composed of a four-color simultaneous camera on a 1.52-m telescope at the Teide Observatory in Tenerife. MuSCAT2 can observe simultaneously in four colours, g (400 to 550 nm), r (550 to 700 nm), i (700 to 820 nm), and zs (820 to 920 nm) bands \citep{Narita2019}. This work includes 97 observations of difficult transits, requiring a larger telescope to be observed (e.g. TOI-2136b, GJ9827d, GJ486b).

\subsection{Data from the ASTEP telescope}

In an effort to make the best use of all resources beyond the data already utilised in \exoclock\, we are open to new collaborations. This work includes 12 light curves from the ASTEP telescope, marking the first steps in a collaboration between the two projects. 

ASTEP (Antarctic Search for Transiting ExoPlanets) is a 40 cm telescope installed at the Concordia station, Dome C, Antarctica that operates during the polar winter from March to September \citep{Fressin2005,Daban2010,Mekarnia2016}. The continuous night and excellent atmospheric conditions make it well suited for high precision time series photometry such as exoplanet transit observations. The telescope was installed in 2010 and upgraded in 2022. The project is a collaboration between Laboratoire Lagrange (CNRS UMR 7293), the University of Birmingham, and the European Space Agency.

\subsection{Data from the Europlanet Telescope Network telescopes}

The Europlanet Telescope Network \citep{Heward2020} provides access to professional and trained amateur astronomers involved in planetary science or exoplanet research to small and medium-sized telescopes from professional observatories in the network around the globe. This work includes 24 light-curves obtained from three telescopes belonging to this network by amateur astronomers who received funding for several nights of telescope time under the Europlanet 2024 RI NA2 Call: the IAC80 telescope at the Teide Observatory (Tenerife), the 1.23m telescope at the Calar Alto Observatory (Almería) and the Joan Oró telescope \citep[TJO,][]{Colome2010} at the Montsec Observatory (Lleida). This includes at least 11 transits of challenging targets were the use of larger apertures is necessary (e.g. TOI-4479b, TOI-1272b, TOI-969b,
K2-284b, LHS1478b), highlighting the importance of the collaboration with such facilities. 

\subsection{Data from the LCO and Telescope Live telescopes}

LCO is an international network of 25 telescopes with diverse sizes (1 m and 40cm) established in several locations \citep[e.g. Australia, Chile, Texas,][]{lco}. The scope of the network is to facilitate scientific, outreach and education projects, while Telescope Live is a network of 10 robotic telescopes \href{https://telescope.live/home}{https://telescope.live/home} with focus on providing data resources for astrophotography. The \exoclock\ project has been awarded a few tens of observing hours from both networks to follow-up exoplanet transits with the collaboration of citizen scientists. 
To facilitate this effort, the \exoclock\ team constructed a dedicated campaign aimed to interested members of the public. The promotion of the campaign was done through \exoclock\ and Ariel media and social media and received over 100 applications including amateur astronomers, members of the general public, educators, school and university students. The successful participants had to complete a series of tasks under the guidance of the \exoclock\ team. In total, four transits were observed by LCO and four by Telescope Live and were analysed by the participants following the observing and analysis techniques described above, to ensure homogeneity in the results.

\subsection{Mid-time points from the literature}

From the available data in the literature, we used only mid-time points that refer to individual transits, while excluding reference mid-time values that represented ephemerides \markchange{(for the references see Table \ref{tab:parameters} in Appendix \ref{sec:sup-info})}. We also excluded mid-time points from space observations that were already integrated through the MAST Archive.

\subsection{Data from the ETD Archive}

In this work we used 181 light-curves from the ETD Archive \citep{etd}, that were already integrated in \exoclock\ DRC, as part of a collaboration between \exoclock\ and ETD that started in 2021. We included data following the quality criteria we have set for all light-curves, as described above, while we excluded observations that had an uncertainty higher than 10 minutes.

\begin{table*}
\centering
\caption{Summary of the successful synchronous observations presented in this work.}
\label{tab:synchronous}
\begin{tabular}{c | c c c c c c c c c}
\hline
Date       & Planet    & $D_\mathrm{min}$ (in) & $E_\mathrm{min}$ (h) & $k$ & $D_i$ (in)  & $RI$  & $QI$ & $SE$ \\ \hline
2022-03-26 & TOI-1298b & 11.17                 & 2.98                 & 13   & 8 - 12           & 3.89  & 3.42 & 87.83\% \\
2022-04-15 & TOI-1789b & 15.33                 & 2.16                 & 14   & 7 - 14           & 2.96  & 2.15 & 72.60\% \\
2022-06-14 & HD191939b & 25.56                 & 2.54                 &  9   & 8 - 14           & 1.39  & 1.21 & 86.91\% \\
2023-04-19 & TOI-1789b & 15.33                 & 2.16                 & 13   & 8 - 16           & 3.30  & 2.38 & 72.35\% \\
2023-12-05 & TOI-942b  & 24.90                 & 2.71                 &  2   & 17               & 1.23  & 0.67 & 54.79\% \\ \hline

\end{tabular}
\end{table*}

\subsection{Synchronous observations}

In \cite{exoclock3}, we estimated that an average telescope in the \exoclock\ network \markchange{-- i.e. a ground-based small- or medium-sized telescope --} can achieve an S/N on the transit of: 

\begin{equation}
\begin{split}
 & S/N = \frac{d}{\sigma_d} = \\ 
 & \frac{0.85 d \sqrt{\pi (D/2)^2 t_e}}{0.135 + 10^{-2.99 + 0.2 R}} \sqrt{\frac{T_{oot}  T_{int}}{(t_e+t_o) (T_{oot} + T_{int})}}
\end{split}
\end{equation}

\noindent where $d$ is the transit depth, $\sigma_d$ is the uncertainty on the transit depth, $D$ is the telescope diameter in inches, $t_e$ is the exposure time in seconds, $t_o$ is the overhead time in seconds, $T_{oot}$ is the observing time out-of-transit in seconds and $T_{int}$ is the observing time in-transit in seconds \markchange{and $R$ is the magnitude of the star in the R band}. 

\markchange{In our observing protocol, we suggest to the observers to observe for one hour before and one hour after the transit, with exposure times at least as long as the overheads (the dead time between exposures), therefore $T_{oot}=7200$ and $t_e=t_o$. Moreover, we can set $T_{int}$ = $t_{14}$ (transit duration) in seconds,  so the equation becomes:}

\begin{equation}
S/N
=\frac{0.85 d D}{0.135 + 10^{-2.99 + 0.2 R}} \sqrt{\frac{900 \pi t_{14}}{(7200 + t_{14})}}
\end{equation}

If we then request the minimum S/N to be 6, we can estimate the minimum telescope diameter $D_{min}$ \markchange{(in inches)} required to observe a transit as:

\begin{equation}
D_\mathrm{min}
= \frac{0.135 + 10^{-2.99 + 0.2 R}}{0.14d} \sqrt{\frac{7200 + t_{14}}{900  \pi  t_{14}}}
\end{equation}

\markchange{\noindent which corresponds to equation C4 of \citet{exoclock3}, with the correction that the factor of 6 appears in the denominator rather than the numerator.} In this case, the minimum total exposure time, $E_\mathrm{min}$, \markchange{in seconds}, will be: 

\begin{equation}
E_\mathrm{min} =  \frac{7200 + t_{14}}{2}
\end{equation}

In this work, we experimented with combining multiple telescopes observing the same transit simultaneously from different locations. Such an approach, if successfully implemented, can enhance the capabilities of the \exoclock\ network and give the small telescopes access to more difficult targets. We attempted a number of such simultaneous campaigns where multiple telescopes having $D \leq D_{min}$ observed in coordination. We present here those campaigns that we considered successful. For a campaign to be considered successful, it had to have:

\begin{equation}
    RI = \sqrt{\frac{  \sum_{i=0}^{i=k} { \left( D_i^2 E_i \right) }}{D_\mathrm{min}^2 E_\mathrm{min}}} > 1
\end{equation}

\noindent where $D_i$ is the telescope size used for each individual observation, $E_i$ is the total exposure time of each individual observation, and $k$ is the number of individual observations. We can define the above quantity as the Resource Index ($RI$) because it gives a measure of how many times more resources (in telescope size and time) were used in each synchronous observations, compared to the minimum required.

The analysis of a synchronous campaign is performed using different trend parameters for each observation (airmass de-trending), different limb-darkening coefficients for each observation (based on the filter used), but forcing the same $R_p/R_s$ and $T_\mathrm{mid}$ for all the observations. This results in $3\times k + 2$ parameters for every synchronous campaign.

Table \ref{tab:synchronous} presents the characteristics and the results for the five synchronous observations that we considered successful. To quantify the performance of the final fit we can examine the uncertainty on the final transit depth, $\sigma_d^\mathrm{OBS}$, with respect to $d/6$ -- i.e. the uncertainty on the final transit depth if the transit S/N was six. We define this as the Quality Index ($QI$):

\begin{equation}
    QI = \frac{d}{6\sigma_d^\mathrm{OBS}}
\end{equation}

Finally, we can examine how effectively the resources used were combined to produce the final results by comparing $RI$ and $QI$. We define this as the Synchronous Efficiency ($SE$):

\begin{equation}
    SE = 100 \frac{QI}{RI}
\end{equation}

\begin{figure}
\centering
\includegraphics[width=\columnwidth]{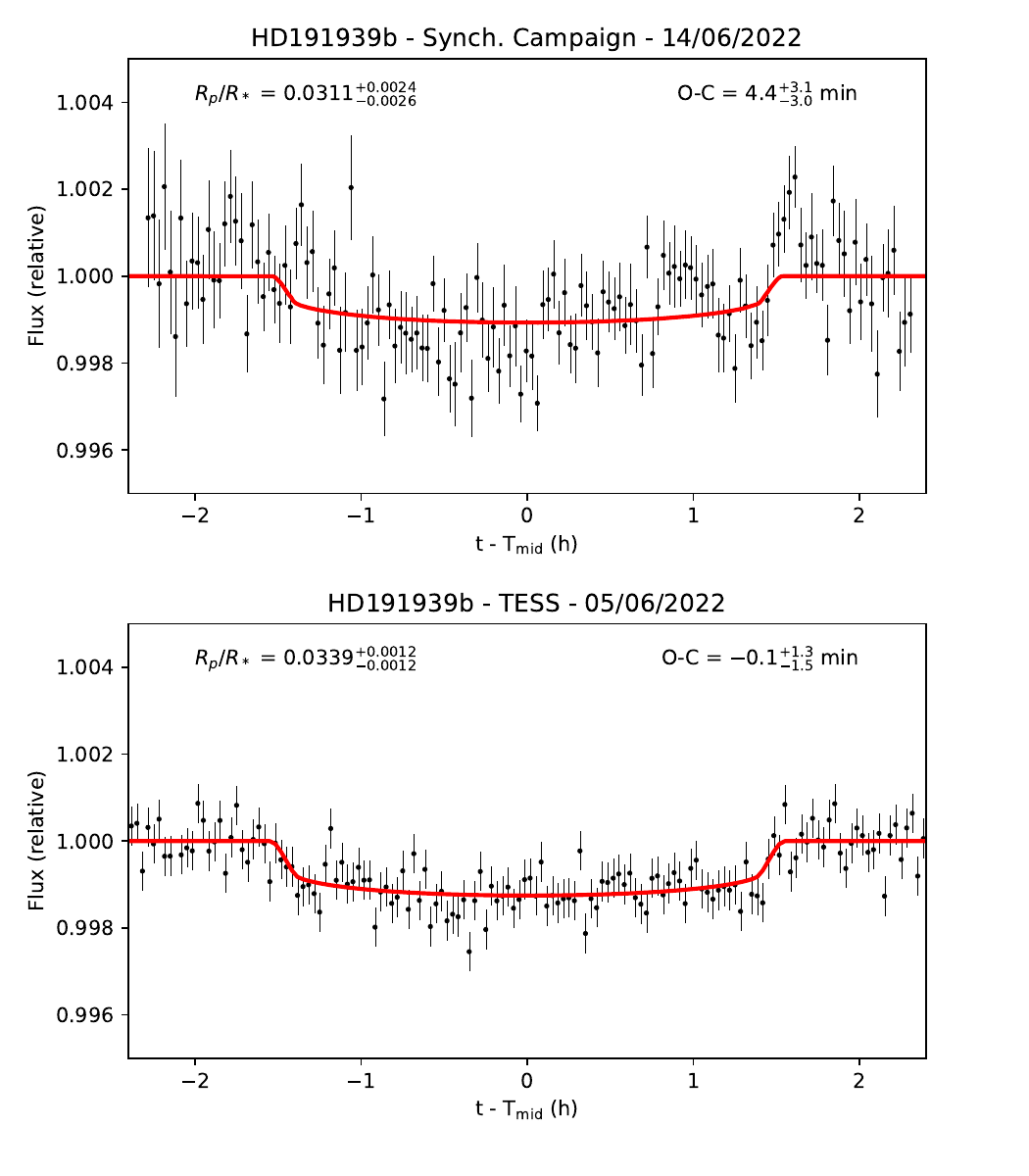}
\caption{De-trended and binned observations for the 2022-06-14 synchronous campaign on HD191939b compared to a TESS observation acquired 1 orbit earlier. Binning has been performed only for visualisation purposes. The fitting has been performed on the original data, without binning.}
\label{fig:HD191939b}
\end{figure}

\begin{figure*}
\centering
\includegraphics[width=0.8\textwidth]{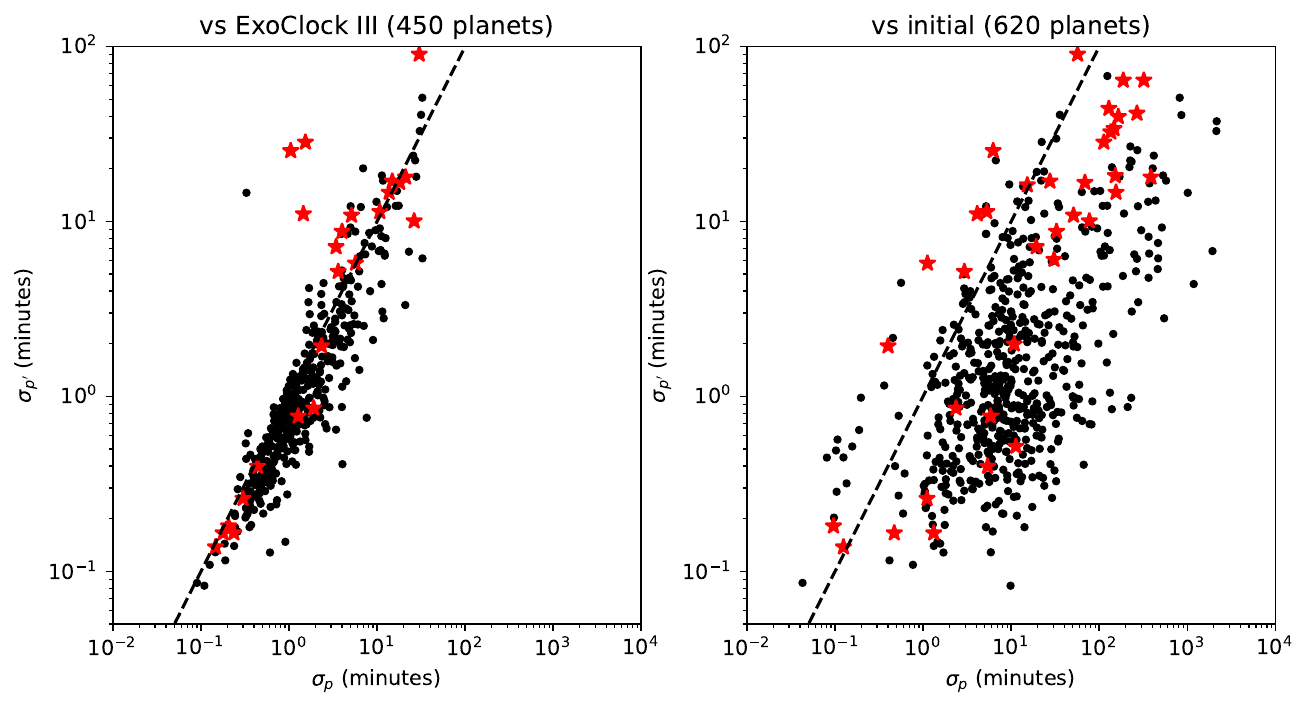}
\caption{Comparison of the 2029-prediction uncertainties between this work and \exoclock\ III (left), or between this work and the ephemerides used when the planet was inserted to the \exoclock\ target list (right). With the red star we indicate the planets with a ``TTVs'' flag. In both  panels, the dashed lines indicate no change ($\sigma_{p'}=\sigma_{p}$).}
\label{fig:before_after}
\end{figure*}

Although the sample is small and we cannot draw statistical conclusions about the behaviour of synchronous observations overall, we can see that this technique can be successful, reaching efficiencies even above 80\% when a large number of observations is combined. We would like to focus the attention of the reader on the case of HD191939b, a planet with a transit depth of 1.3 mmag around a star of $R_\mathrm{mag}=8.5$ (Fig. \ref{fig:HD191939b}) that would \markchange{require} a telescope larger than 25 inches to be detectable. The combination of nine observations with telescopes smaller than 14 inches, summing approximately 27 observing hours, resulted in a strong detection with $S/N = 6.27$ and $\sigma_{T_\mathrm{mid}}=3.1$ minutes, comparable to the capabilities of TESS ($S/N = 13$ and $\sigma_{T_\mathrm{mid}}=1.5$ minutes). Such shallow transits around bright stars are difficult to observe even with large telescopes, due to saturation. Therefore, synchronous observations with small telescopes may be our only window to follow-up such targets.

\section{Results}

\subsection{Ephemerides}

In this work, we provide a homogenous list with updated ephemerides for 620 of the total 775 planets that are currently in the \exoclock\ target list\footnote{\href{https://www.exoclock.space/database/planets}{https://www.exoclock.space/database/planets}}. We integrated all data from the resources described above (ground-based telescopes, space observations and mid-time values from the literature). After calculating the updated zero-epoch point to the weighted average of the available epochs, we fitted a line on the epoch versus the transit mid-time data. For the fitting, we used the MCMC algorithm in the emcee package \citep{ForemanMackey2013}. Following an initial fit, we scaled-up the uncertainties by \markchange{multiplying them with} the RMS of the normalized residuals, to consider excess noise. We performed a new fitting after scaling-up and Table \ref{tab:updated_ephemerides} shows the new ephemerides for all the planets and the references to the literature mid-time values used.

\begin{table}
\centering
\caption{Categories of ephemerides in comparison with the previous \exoclock\ publication and the values at the beginning of the project.}
\label{tab:before_after}
\begin{tabular}{l | c c c}
\hline
					&			& ExoClock IV vs 	&		\\
					& ExoClock III 	& ExoClock II 		& Initial	\\ \hline
planets   				& 450  	        & 180        		& 620	\\
Sign. improved        		& 3.1\%		& 0.0\%			& 32.6\%	\\ 
Drifting				& 1.1\%		& 1.1\%			& 11.9\%	\\ 
Improved				& 14.7\%		& 40.0\%			& 38.4\%	\\ 
No change			& 70.9\%		& 55.6\%			& 9.2\%	\\
TTVs				& 6.2\%		& 3.3\%			& 6.8\%	\\
Worse				& 4.0\%		& 0.0\%			& 1.1\%	\\ \hline
\end{tabular}
\end{table}

Figure \ref{fig:before_after} shows the uncertainties in the 2029-predictions before and after the updates presented in this work ($\sigma_p$ and $\sigma_{p'}$, respectively), \markchange{where $p$ and $p'$ denote the corresponding predicted mid-times}, while Table \ref{tab:before_after} lists six categories of the ephemerides status. ``Significantly improved'' refers to the ephemerides that were giving 2029-predictions with uncertainties greater than the target uncertainty of 1/12$^\mathrm{th}$ of the transit duration, $t_{14}$, ($\sigma_p>t_{14}/12$) as described in \cite{exoclock1}. The term ``drifting'' refers to ephemerides with 2029-predictions that drifted more than the target uncertainty ($|p-p'|>t_{14}/12$). From the remaining ephemerides, the term ``Improved'' refers to those ephemerides for which the 2029-prediction uncertainties have been improved by more than one minute ($\sigma_{p'}<\sigma_p - 1$), ``Worse'' refers to those ephemerides for which the 2029-prediction uncertainties are now worse by more than one minute ($\sigma_{p'}>\sigma_p + 1$), while ``No change'' refers to those ephemerides for which the 2029-prediction uncertainties have not changed by more than one minute ($|\sigma_{p'} - \sigma_p| < 1$). Finally, the ``TTVs'' flag refers to ephemerides that deviate from linear behaviour \markchange{(see the following section)}.

Although not expected, we found that a number of ephemerides (37) were significantly improved, or drifting, or worse than \markchange{the} previous \exoclock\ release. Most of these cases refer to multi-planetary systems so the variability could be related to planet-planet interactions which, however, do not show a statistically significant deviation from a linear ephemeris to be flagged as ``TTVs''. The rest of the cases (EPIC211945201b, GJ1252b, HD219666b, K2-115b, K2-116b, K2-132b, K2-334b, LHS3844b, TOI-1201b, TOI-1478b, TOI-169b, TOI-640b, TOI-892b, WASP-169b, WASP-68b) are related to low time-coverage in the previous release. As noted in \cite{exoclock3} our previous sample was not completely bias-free and this is the reason we see here a small percentage of problematic ephemerides, which points towards the significance of increasing the time-coverage to have a bias-free catalogue.

When comparing the results of this work with the initial ephemerides, we can see that on average the ephemerides have improved by approximately one order of magnitude (the median improvement in the 2029-prediction uncertainty is 7.9 times). Approximately 45\% of the ephemerides (those that are significantly improved, or drifting) were in need of an update to avoid an impact on the final Ariel schedule. If we add to this percentage the planets that are affected by TTVs, then we conclude that 50\% of the planets added to our target list at any time need to be followed-up.   This result is similar to the previous \exoclock\ release, indicating that the newly discovered TESS planets follow a similar statistical behaviour with the planets that were part of our initial target list in terms of uncertainties and bias in their ephemerides.

\begin{table*}
\centering
\caption{Planets not in multi-planetary systems identified with deviations from a linear ephemeris. Long and Short refer to long- or short-term variations. Q refers to the quadratic term of the ephemeris.}
\label{tab:ttvs}
\begin{tabular}{c | c c c c}
\hline
Planet		& points	& variations after linear fit  		& Q 		& variations after quadratic fit \\ \hline
HAT-P-7b & 688 & Short \& Long & $74.6_{-4.2}^{+4.1}\times10^{-11}$ & None\\
HD332231b & 5 & Short \& Long & $-16.1_{-1.8}^{+1.8}\times10^{-6}$ & None\\
KELT-9b & 78 & Short & $-0.4_{-2.2}^{+2.1}\times10^{-10}$ & Short\\
TOI-201b & 14 & Short \& Long & $3.1_{-1.1}^{+1.1}\times10^{-5}$ & Short \& Long\\
TOI-1333b & 12 & Short & $2.8_{-4.1}^{+4.0}\times10^{-8}$ & None\\
TrES-3b & 383 & Short \& Long & $-11.0_{-2.1}^{+2.1}\times10^{-11}$ & Short \& Long\\
WASP-4b & 150 & Long & $-9.6_{-1.3}^{+1.4}\times10^{-11}$ & None\\ 
WASP-12b & 411 & Short \& Long & $-53.7_{-1.3}^{+1.2}\times10^{-11}$ & None\\
WASP-33b & 69 & Short & $18.1_{-6.0}^{+6.1}\times10^{-11}$ & None\\
WASP-19b & 218 & Short \& Long & $-63.6_{-8.2}^{+8.7}\times10^{-12}$ & Short\\
WASP-135b & 111 & Short \& Long & $4.9_{-2.3}^{+2.3}\times10^{-10}$ & Short \& Long\\
WASP-161b & 9 & Short \& Long & $-578.1_{-6.4}^{+6.4}\times10^{-9}$ & None\\ \hline

\end{tabular}
\end{table*}

\subsection{Deviations from linear ephemerides}

For all the 620 planets studied in this work, we studied the possibility of non-linear ephemerides. Identifying these cases is important for Ariel, in order to implement different ephemerides and produce precise transit and eclipse time predictions. Such deviations occur as a result of physical processes, like stellar activity, orbital decay, orbital precession, and planet-planet interactions in multi-planetary systems \citep{Agol2018}.

To identify non-linear behaviour in transit timings we used the non-normalised Lomb-Scargle periodogram on the residuals of the linear ephemeris fit, as implemented in the python package SciPy \citep{Lomb1976, Scargle1982, Virtanen2020}, similarly to \cite{exoclock3}.

\begin{figure*}
\centering
\includegraphics[width=\textwidth]{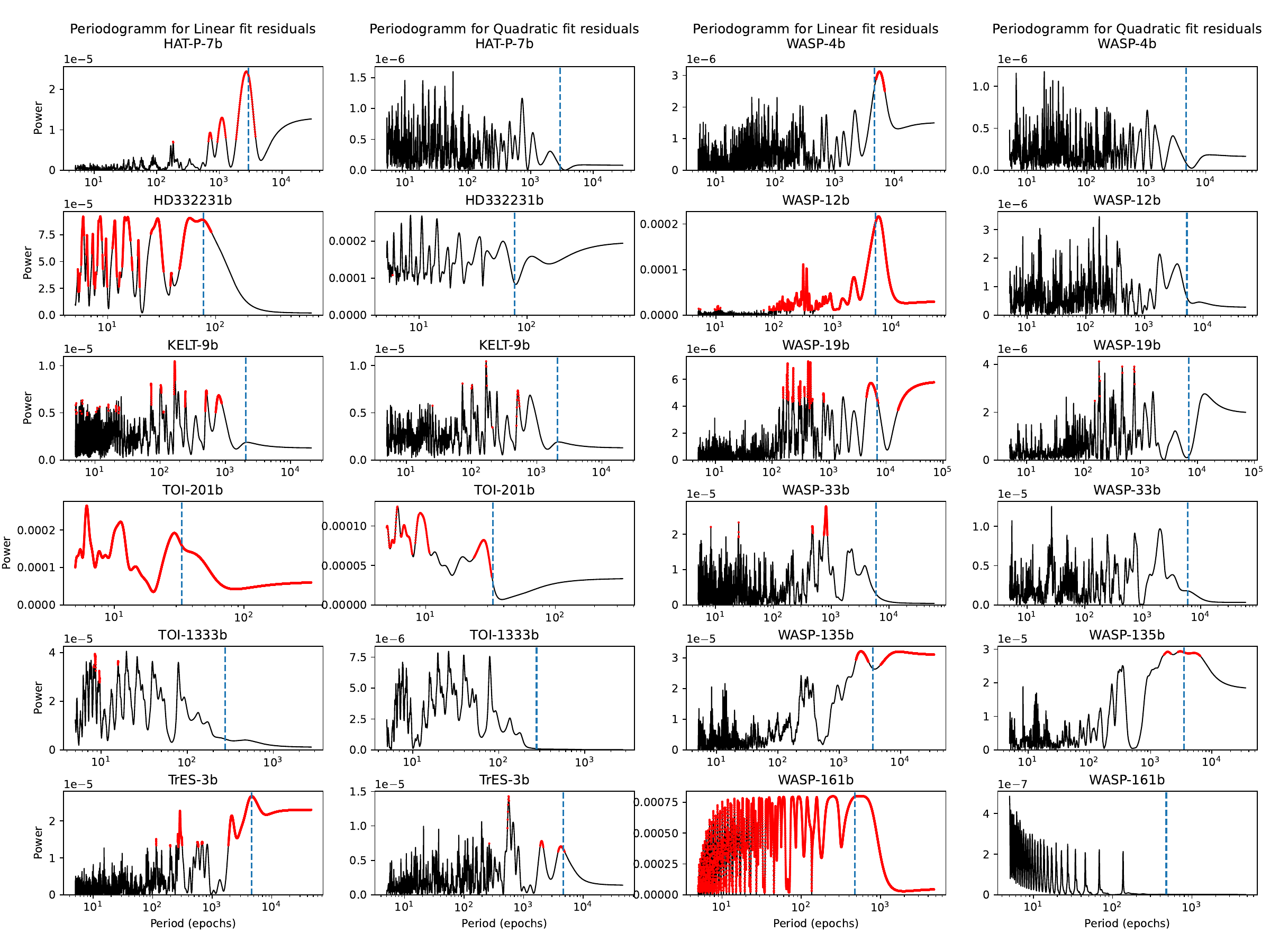}
\caption{Periodogramms for the fitting residuals (linear and quadratic) for the 12 planets with TTVs but without transiting companions. The red parts indicate periods with FAP lower than 0.13\% and the vertical line indicates the total time span of the data used.}
\label{fig:periodogramms}
\end{figure*}

We first calculated the power of the periodogram of the residuals of the linear ephemeris fit for periods between 5 epochs and 10 times the full time-span of the observations. Then we produced 100,000 periodograms from time-series that had the same epochs as the residuals but the mid-time values were drawn from a normal distribution of zero mean and STD equal to the uncertainty of each observed data point (we name these periodograms Pa). Finally, we produced 100,000 periodograms from time-series that were equal to the residuals plus a value drawn from a normal distribution of zero mean and STD equal to the uncertainty of each observed data point (we name these periodograms Pb).

The FAP for each period was then defined as the percentage of Pb periodograms that had greater power than the 99.87\% (3$\sigma$) upper limit of the Pa periodograms. The ``TTVs'' flag was then attributed to those planets with periodogram peaks that has FAPs lower than 0.13 (3$\sigma$). Detected variations were categorised as short-term or long-term, based on the time span of all available data. Long-term are the variations that are longer than 90\% of the total time-span of the data used.

We found 42 planets with statistically significant TTVs. 30 of these planets belong to multi-planetary systems, that can explain those TTVs (namely: HD106315c, HD108236b, HD191939c, HD191939d, HD28109c, K2-19b, K2-19c, K2-21c, KOI-12b, KOI-94c, Kepler-18d, Kepler-396c, L98-59b, L98-59c, TOI-1130b, TOI-1130c, TOI-1136c, TOI-1136d, TOI-1136e, TOI-1136f, TOI-1246d, TOI-125b, TOI-2076b, TOI-2076c, TOI-216.01, TOI-216.02, TOI-270c, TOI-270d, TOI-712c, WASP-148b). For the remaining 12 planets shown in Table \ref{tab:ttvs} and Figure \ref{fig:periodogramms} (namely: HAT-P-7b, HD332231b, KELT-9b, TOI-201b, TOI-1333b, TrES-3b, WASP-4b, WASP-12b, WASP-19b, WASP-33b, WASP-135b, WASP-161b) we performed extra analysis assuming a quadratic ephemeris. In this group, we found that the quadratic terms were statistically significant (3$\sigma$) for HAT-P-7b, HD332231b, TrES-3b, WASP-4b, WASP-12b, WASP-19b, and WASP-161b. A discussion on each planet is included in Section \ref{sec:discussion}.

\section{Data release D}

The \markchange{fourth} data release of the \exoclock\ project (DRD) includes two data products: the Catalogue of Observations (\exoclock, ETD, space observations), and the catalogue of \exoclock\ ephemerides. \markchange{All data products and their descriptions can be found through DOI: 10.17605/OSF.IO/WPJTN, hosted by the Open Science Framework.}

\subsection{Catalogue of Observations}

The Catalogue of Observations contains all the light-curves and literature mid-time points summarised in Table \ref{tab:data}. In the online repository, each light-curve is accompanied by:
\begin{enumerate}
\item metadata regarding the planet, the source, the observation, the instrument, and the data format;
\item the pre-detrended light curve, filtered for outliers, converted to BJD$_\mathrm{TDB}$ and flux formats, with scaled uncertainties;
\item the fitting results, including the de-trending method used and its parameters;
\item the de-trended light curve, enhanced with the de-trending model, the transit model and the residuals;
\item fitting diagnostics on the residuals.
\end{enumerate}

\subsection{Catalogue of \exoclock\ Ephemerides} 

The new catalogue of \exoclock\ ephemerides contains the updated ephemerides for the 620 planets studied in this work (see also Table \ref{tab:updated_ephemerides}), accompanied by metadata regarding the planet, and flags concerning the detection of TTVs.

\section{Discussion} \label{sec:discussion}

\subsection{Follow-up efficiency and comparison with previous ExoClock data releases}

To evaluate the efficiency of our follow-up strategy we compared the newly acquired data by TESS (after 1/1/2022) to the predictions of the ephemerides published in \exoclock\ DRC \citep{exoclock3} and DRB \citep{exoclock2}. We decided to use TESS data only as a calibrator because, like Ariel, the TESS observations include long base-lines before and after the transit, while ground-based observations have limited base-lines (usually one hour before and after the transit). The results are presented in Table \ref{tab:drift2}, where we can see that the percentage of measurements with $\frac{O-C}{\sigma_{O-C}} < 1$ when compared with the ephemerides published in \exoclock\ DRC is 62.32\%, indicating that the uncertainties in our previously released ephemerides were almost following the normal distribution at the $1\sigma$ level. The success rate at $3\sigma$ level,  -- i.e. the percentage of measurements with $\frac{O-C}{\sigma_{O-C}} < 3$ -- for DRC is now at 98.34\%, gradually approaching the normal distribution. These results underline the fact that the strategy followed in the \exoclock\ project is efficient and capable of producing a consistent catalogue of reliable ephemerides as the time-span of observations is increasing.  More over, the above indicate that for the full Ariel candidate target list, the percentage of problematic Ariel observations (in terms of timing) will be below 2.0\%. As the time span of the follow-up campaigns increases, this percentage is expected to decrease even more.

\begin{table}
\centering
\caption{Success rate of each ephemeris set on the TESS data acquired after 1/1/2022. For DRD this calculation is optimistic (the TESS data acquired after 1/1/2022 were included in the production of DRD), but it is shown for completeness.}
\label{tab:drift2}
\begin{tabular}{l | c c c c}
\hline
vs			                    	& DRD	& DRC     & DRB     & Initial \\ \hline
planets					& 424	& 308 	& 130 	  & 424     \\
measurements   			& 4716	& 3604 	& 1585 	  & 4716    \\
$\frac{O-C}{\sigma_{O-C}} < 1$	& 66.39\%	& 62.32\% & 48.01\% & 53.71\% \\
$\frac{O-C}{\sigma_{O-C}} < 2$	& 93.70\%	& 91.32\% & 76.53\% & 82.78\% \\
$\frac{O-C}{\sigma_{O-C}} < 3$	& 99.28\%	& 98.34\% & 93.50\% & 92.88\% \\
$\frac{O-C}{\sigma_{O-C}} < 4$	& 99.96\%	& 99.61\% & 98.30\% & 95.50\% \\ \hline
\end{tabular}
\end{table}

\subsection{New needs in the project}

TESS targets are quite challenging for several reasons; the particularity of the TESS targets implies special requirements for follow-up observations. For example, large ground-based telescopes cannot easily observe the brightest of these targets due to the scintillation noise and the risk of saturation. On the other hand, some of the TESS transits have shallow depths which means that small sized telescopes are insufficient to detect their signal. Many planets are also part of multi-planetary systems, resulting in TTVs and therefore need special attention for continuous monitoring. 

\begin{figure}
\centering
\includegraphics[width=\columnwidth]{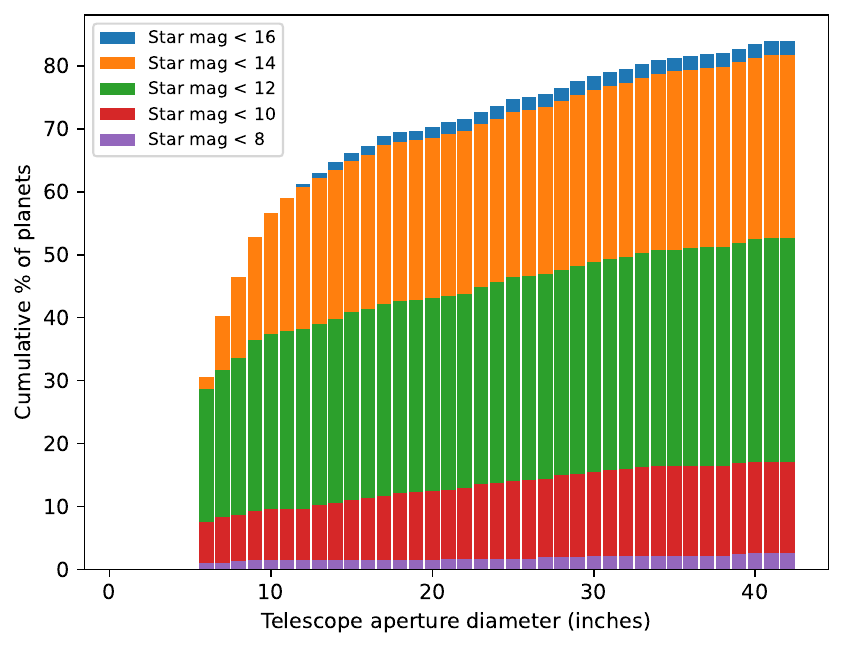}
\caption{Distribution of available planets per magnitude and telescope aperture diameter.}
\label{fig:network_capabilties}
\end{figure}

With the increased number of TESS planets in the Ariel target list, it becomes apparent that our current network of telescopes is not enough to efficiently correspond to the new needs of the project. Figure \ref{fig:network_capabilties} shows the updated capabilities of the telescopes in the \exoclock\ network. This plot has been produced assuming that the minimum telescope diameter $D_{min}$ required to observe a transit with depth $d$ and duration $t_{14}$ in seconds around a star of magnitude $R$ in the red filter is (see \cite{exoclock3}):

\begin{equation}
D_{min}
=  \frac{0.135 + 10^{-2.99 + 0.2 R}}{0.14  d } \sqrt{\frac{7200 + t_{14}}{900  \pi  t_{14}}}
\end{equation}

In \cite{exoclock3}, the percentage of planets in the \exoclock\ target list that could be followed-up by a 16-inch telescope was 75\%, while now this percentage is 67\%, with the expectation that it will reduce further as more TESS discoveries are integrated. For this reason, new strategies and methods are required and we have already paved the way for the new era of transit monitoring. 

The integration of data from larger facilities and telescopes located in sites beyond the \exoclock\ network (like MuSCAT2 and ASTEP) enabled the extended coverage of planets for which we did not have available observations. We plan to continue the synergies we initiate with this work to increase the coverage of planets. With the simultaneous observations is clear that we can achieve a high photometric precision in following-up planets with low S/N around bright stars, and we plan to organise more such efforts in the future. Finally, new observations from space telescopes (CHEOPS, JWST and others) will facilitate these efforts.

We also plan to include planets from the TOI candidate list as these might be interesting for characterisation studies by Ariel. The uncertainties in the ephemerides of these planets are increasing while we are waiting for their confirmation and therefore their timings might be completely lost by the time they get confirmed.
\citep{Hord2024}.

\subsection{Need for continuous monitoring}

As demonstrated in Table \ref{tab:before_after}, around 45\% of the initial planet ephemerides have large uncertainties or drifts (the categories marked as "significantly improved" or "drifting"). This percentage signifies that a considerable number of the planet ephemerides require update in order to construct an efficient observing plan by the time Ariel flies, and therefore continuous monitoring is essential.

Moreover, we notice that the new ephemerides set is not completely bias-free, despite that biases have been reduced (see comparison with DRD in Table \ref{tab:drift2}). Therefore, it is essential to extend the coverage for all planets in our list by collecting new data from different resources and extending the time-span of the follow-up observations.

Finally, continuous monitoring is necessary for the planets flagged as ``TTVs`` in order to construct a precise ephemeris that includes the dynamics of these systems.

\subsection{TTVs signals}

\subsubsection{Comparison with ExoClock DRC}

In agreement with \cite{exoclock3}, here we find that HAT-P-7b, TrES-3b, WASP-4b, WASP-12b, and WASP-19b show consistent long-term TTVs. These are indicated by the peaks in the periodograms that have periods similar to the full time-span of the data, which, however, disappear after fitting for a quadratic ephemeris (Table \ref{tab:ttvs}). Other studies have also suggested the presence of quadratic trends in these planets (e.g. \cite{Narita2012} for HAT-P-7b, \cite{2020AJ....160...47M} for TrES-3b, \cite{Baluev2020} for WASP-4b, \cite{2020ApJ...888L...5Y} for WASP-12b, \cite{2023Univ...10...12K} for WASP-19b).  

On the contrary, Qatar-1b and WASP-56b no longer show significant deviations from a linear ephemeris. For both planets we acquired a large number of new data, after flagging them as ``TTVs'' in the previous data release, indicating that the signals found in the previous analysis could be driven by biases in the older observations. Both cases will be closely monitored in the future. 

\subsubsection{New planets with ``TTVs'' flag}

\paragraph{HD332231b}

So far, it has not been clear whether the system of HD332231b exhibits TTVs. Spectroscopic observations did not yield enough evidence suggesting a companion planet, and the same conclusion was deduced by radial velocity measurements \citep{2020AJ....159..241D}. It was speculated though, that a slight linear trend in those time series and residuals might indicate the presence of an outer companion \citep{2020AJ....159..241D} but follow-up observations by \citep{Sedaghati2022} did not reveal any statistically significant signal for a detection. 

Despite the small numbers, the new observations from TESS and one light curve  from the \exoclock\ network show significant variability, and a significant quadratic term in the ephemeris of the planet. To verify the origin of these TTVs, further observations are required.

\paragraph{KELT-9b}

An early analysis of KELT-9b’s orbit used transit light curves and radial velocity analyses to constrain the model to either a fiducial one or one with TTVs \citep{2017Natur.546..514G}. However, it was found that the TTV model and the fiducial model had nearly-identical uncertainties associated with them, so the fiducial one was adopted (ibid.). A study by \citep{2022ApJS..259...62I} found that TTVs were not present within this system. \markchange{ \cite{2023AnA...669A.124H} fitted different models to determine the most appropriate explanation for the timing deviations on KELT-9b. Their result suggest that apsidal  precession with a non-zero eccentricity can better describe the deviations, however, orbital decay or even a combination of the two models can be a possible solution. Moreover, it is speculated that the eccentricity could derive from the migration history of the planet or from the presence of a third, as-yet unseen body in the system (ibid.).} Our analysis of 78 data points shows short-term variations but further observations are necessary to differentiate the various models and the explanations for the deviations.

\paragraph{TOI-201b}

TOI-201b is a warm giant planet orbiting an F-type star that exhibits long-term variability due to stellar activity \cite{2021AJ....161..235H}. \markchange{The analysis of space data points and two ExoClock observations shows both short and long term variations. \cite{2025ApJ...988L..63M} performed a TTV analysis in combination with radial-velocity data and a recent transit from TESS to report the detection of TOI-201c, a long period giant planet. The new planet is the most possible explanation for the short-term variations identified in our analysis, but more data are required to explore the long-term variability.}

\paragraph{TOI-1333b}

TOI-1333b orbits a subgiant star, while two additional light sources are depicted in image processing \citep{2021AJ....161..194R}. It is suggested that the light curves of TOI-1333b are diluted due to the light emitted from these other stars \citep{2021AJ....161..194R}. The furthest star was identified as a chance alignment, while the closest one was identified as a companion star \citep{2021AJ....161..194R}. In our study we used 12 points from TESS observations and we found short-term variations from the linear ephemeris. Although \markchange{the companion star could be the source of the TTVs, such interaction should result to variations at longer time-scales and the small number of data points does not allow for their detection.}

\paragraph{WASP-33b}

WASP-33b orbits around a $\delta$ Scuti variable star \citep{2010MNRAS.407..507C} and with a short orbital period  \citep{2020AnA...639A..34V}, is expected to be affected by heating, stellar winds and tidal forces from its host star. It is clear that WASP-33b interacts intensively with its star, and some of the possible interactions include mass transfer \citep{Kovacs2013}. Other types of interactions that affect the orbit of the planet have been suggested, but were excluded, such as the existence of additional bodies, low spherical distortions due to the $\delta$ Scutti pulsations and others \citep{Kovacs2013}. In 2018, McDonald et al. observed a slight orbital expansion but stated that this cannot be accounted as significant detection of TTVs. Due to spin-orbit misalignment \citep{2010MNRAS.407..507C}, non-radial changes to the orbit of WASP-33b can be expected \citep{McDonald2018}, and more complex TTV signals could appear over long periods. \markchange{We identified short-term TTVs which can result from the stellar variability which is introducing bias to the timing measurements. In addition, new observations from CHEOPS confirm nodal precession \citep{2025AnA...693A.128S} which can cause the variations with time-scales around 1000 epochs.}

\paragraph{WASP-135b}

WASP-135b is a hot Jupiter orbiting a Sun-like star discovered in 2016 by \citet{2016PASP..128b4401S}. WASP-135b receives high levels of insolation due to the proximity of the planet to its star, has an inflated radius and shows evidence of a transfer of angular momentum from the planet to its host star \citep{2016PASP..128b4401S}. Until now, it has not been evident whether the system displays TTVs. The last photometric analysis of WASP-135b \citep{Ozturk2021} suggests the possibility of a decrease in its orbital period. However, confirming this hypothesis requires obtaining new mid-transit times. In addition, there is an age difference between the isochronal and gyrochronological age of the star that may indicate stellar spin-up \citep{2016PASP..128b4401S}, although this hypothesis is weak.

\paragraph{WASP-161b}

 Significant TTVs have previously been detected in WASP-161b using TESS and archival data, with shifts in the transit midpoints observed in January 2019 and 2021 \citep{2019AJ....157...43B}, diverging from previous ephemerides by approximately 67 minutes and 203 minutes. These TTVs align with a quadratic model, indicating a constant period derivative quantified as -1.16$\times$ 10$^{-7}$ days per day, suggesting possible tidal dissipation and a decaying orbital period \citep{YangChary2022, Yang2022}. The largest TESS timing offset noted was -203.7 $\pm$ 4.1 minutes \citep{Shan2023}. Explanations such as period decay and apsidal precession have been proposed, but inconsistencies remain, which need to be supplemented with additional data to find a clearer cause for the origin of TTVs \citep{YangChary2022, Yang2022, Shan2023}. Our analysis demonstrated short and long term variations and we plan to monitor these with further data.

\section{Conclusion}

The ExoClock project has been continuously operating for the past 6 years following open science strategies during all stages: open software, hardware, data, and open to contributions from diverse communities including academics and non-academics such as citizen scientists and school students.This work presents the updated ephemerides for 620 planets which are current candidates of the Ariel Mission Reference Sample. After comparison of the new catalogue with the previous version it is shown that biases are reduced, which underlines that the approach of the ExoClock project is efficient for generating reliable ephemerides. Our study demonstrated that 45\% of the planets required an update, a result that highlights the need for continuous monitoring. The new catalogue includes the updated ephemerides for a large sample of TESS planets which are challenging for observations due to shallow transits or bright host stars. The new data from larger telescopes and sites beyond the usual ExoClock network enabled the coverage of planets with lower S/N ratio and planets inaccessible from usual sites of the ExoClock network.
The open science approach of the project has demonstrated to be the most successful way to provide a validated catalogue of planet ephemerides for the Ariel mission. Through the ExoClock project, not only a reliable scheduler for Ariel established but also further collaborations and research efforts are facilitated. These include testing new methodologies such as the synchronous observations and investigating new research ideas, for example planets with TTVs. We plan to continue fostering synergies with large facilities and space telescopes but also monitoring planets with TTVs and conducting further experimental efforts with synchronous observations. This approach facilitates our effort to correspond to the new needs of the project while it accelerates collaborations and progress in the field of exoplanet research.

\section*{Software and Data} 

Software used: Django, PyLightcurve \citep{Tsiaras2016B2016ApJ...820...99T}, ExoTETHyS \citep{Morello2020, exotethys_joss}, Astropy \citep{AstropyCollaboration2013}, emcee \citep{ForemanMackey2013}, Matplotlib \citep{Hunter2007}, Numpy \citep{Harris2020}, SciPy \citep{Virtanen2020}.

\markchange{All the data products can found through the OSF repository with DOI: 10.17605/OSF.IO/WPJTN, alongside their descriptions.}

\section*{Acknowledgements}

The ExoClock project has received funding from the STFC grants ST/W006960/, ST/X002616/ and ST/Y001508/1.

This work has made use of data collected with the TESS mission, obtained from the MAST data archive at the Space Telescope Science Institute (STScI). Funding for the TESS mission is provided by the NASA Explorer Program. STScI is operated by the Association of Universities for Research in Astronomy, Inc., under NASA contract NAS 5–26555.

This work has made use of observations made by the LCOGT network, as part of the LCOGT Global Sky Partners projects ``ExoClock'' (PI: A. Kokori) and ``ORBYTS: Refining Exoplanet Ephemerides'' (PI: B. Edwards).

This work makes use of observations from the ASTEP telescope. ASTEP has benefited from the support of the French and Italian polar agencies IPEV and PNRA, and from INSU, the European Space Agency (ESA) through the Science Faculty of the European Space Research and Technology Centre (ESTEC), the University of Birmingham, the European Union's Horizon 2020 research and innovation programme (grants agreements n$^{\circ}$ 803193/BEBOP), the Science and Technology Facilities Council (STFC; grant n$^\circ$ ST/S00193X/1), the laboratoire Lagrange (CNRS UMR 7293) and the Universit\'e C\^ote d'Azur through Idex UCAJEDI (ANR-15-IDEX-01).

We would like to acknowledge the support provided by the administrators, designers, and developers of the ETD project and of the Czech Astronomical Society both to the ExoClock project but also to the efforts of the whole amateur community through its 10+ years of operation.

Several observations were supported by the Europlanet 2024 RI project that has received funding from the European Union’s Horizon 2020 research and innovation program under grant agreement No 871149. 

This work includes observations made with: the MuSCAT2 instrument mounted in the Telescopio Carlos Sánchez at Teide Observatory; the Joan Oró Telescope (TJO) of the Montsec Observatory (OdM), owned by the Catalan Government and operated by the Institute of Space Studies of Catalonia (IEEC); the IAC80 telescope operated on the island of Tenerife by the Instituto Astrofísico de Canarias in the Spanish Observatorio del Teide; the Centro Astronómico Hispano en Andalucía (CAHA) at Calar Alto, operated jointly by the Junta de Andalucía and the Instituto de Astrofísica de Andalucía (CSIC); the Observatorio do Pico dos Dias/LNA (Brazil); the Madrona Peak Observatory, owned by the non-profit Mark and Candace Williams Family Foundation, dedicated to science education.

This work has made use of observations made by the MicroObservatory which is maintained and operated as an educational service by the Center for Astrophysics, Harvard \& Smithsonian as a project of NASA's Universe of Learning, supported by NASA Award \# NNX16AC65A.

A.B. acknowledges financial support from the Italian Space Agency (ASI) with Ariel grant n. 2021.5.HH.0.

Members from Silesian University of Technology were responsible for (1) observations planning, (2) automation of observatories work, and (3) processing of data from SUTO network. P.J.-W. acknowledges support from grant BKM-2025. Other authors from Silesian University of Technology acknowledge grant BK-2025.

R.A.A. would like to thank Mercedes Gómez, Director of the Observatorio Astronómico Córdoba, for supporting the use of the facilities at the Estación Astrofísica Bosque Alegre (EABA), and Luis Tapia Portillo, Telescope operator at EABA, for his collaboration during the observing nights.

VJSB acknowledges financial support from grant PID2022-137241NB-C41 funded by Agencia Estatal de Investigaci\'on of the  Ministerio de Ciencia, Innovaci\'on y Universidades (MICIU/AEI/10.13039/501100011033) and ERDF/EU.

AAB acknowledge support from M.V. Lomonosov Moscow State University Program of Development.

M.L.C., L.N.C. and R.A.P. acknowledge financial support from the Coordenação de Aperfeiçoamento de Pessoal de Nível Superior -- Brasil (CAPES) -- Finance Code 001 and the Fundo Paraná -- Finance Code 31/24 and also would like to thank the Laboratório Nacional de Astrofísica -- Brasil for the observing time granted at the Pico dos Dias Observatory.

A.D. has made use of the OARPAF telescope within the scientific mission of University of Genova and would like to thank Silvano Tosi, Davide Ricci and Lorenzo Cabona for their support. 

A.M. acknowledges the use of data obtained at the Observatório do Pico dos Dias / LNA (Brazil), Observatório Astronômico – Universidade Estadual de Ponta Grossa (Brazil) and financial support from Conselho Nacional de Desenvolvimento Científico e Tecnológico (CNPq)

T.E has received funding from the European Union's Horizon Europe research and innovation programme under grant agreement No. 101079231 (EXOHOST), and from the United Kingdom Research and Innovation (UKRI) Horizon Europe Guarantee Scheme (grant number 10051045). This work has made use of the ground-based research infrastructure of Tartu Observatory, funded through the projects TT8 (Estonian Research Council) and KosEST (EU Regional Development Fund).

E. E.-B. acknowledges financial support from the European Union and the State Agency of Investigation of the Spanish Ministry of Science and Innovation (MICINN) under the grant PRE2020-093107 of the Pre-Doc Program for the Training of Doctors (FPI-SO) through FSE funds.

A.F. is partly supported by JSPS KAKENHI Grant Numbers JP24K00689 and JP24H00017.

P.G. is supported by the Slovak Research and Development Agency under contract No. APVV-20-0148, internal grant VVGS-2023-2784 of the Faculty of Science, P. J. Šafárik University in Košice and funded by the EU NextGenerationEU through the Recovery and Resilience Plan for Slovakia under the project No. 09I03-03-V05-00008.

E.G. acknowledges financial support from the Gordon and Betty Moore Foundation for the Global Sky Partners program.

G.F.G. would like to thank Andressa Wille (UTFPR), Elvis Mello-Terencio (UTFPR), Giovanni Tauil (UTFPR), Micheli Moura (UFRGS), Nicholas Souza (UFPR), Richards Albuquerque (UTFPR), Vinicius Fochesatto (UTFPR) and Rubens Machado (UTFPR).

J.K. acknowledges support from the Swiss National Science Foundation under grant number TMSGI2\_211697, and from the Swiss NCCR PlanetS. Their work has been carried out within the framework of the NCCR PlanetS supported by the Swiss National Science Foundation under grants 51NF40182901 and 51NF40205606. 

SPL acknowledges the grants from the National Science and Technology Council of Taiwan under project numbers 109-2112-M-007-010-MY3, 112-2112-M-007-011, and 113-2112-M-007-004.

Y.L. would like to thank K.P. Huang (National Museum of Nature Science) for their technical support.

J.L. is partly supported by Astrobiology Center SATELLITE Research project AB022006 and JSPS KAKENHI Grant Number JP24H00017.

S.L. acknowledges financial support from Orizzonti di Lora Stefano

G. M. acknowledges financial support from the Severo Ochoa grant CEX2021-001131-S and from the Ram\'on y Cajal grant RYC2022-037854-I funded by MCIN/AEI/ 10.13039/501100011033 and FSE+.

M.M. is partly supported by JSPS Grant-in-Aid for JSPS Fellows Grant Number JP24KJ0241, JSPS KAKENHI Grant Number JP24K17083.

F. M. acknowledges the financial support from the Agencia Estatal de Investigaci\'{o}n del Ministerio de Ciencia, Innovaci\'{o}n y Universidades (MCIU/AEI) through grant PID2023-152906NA-I00.

N.N. is partly supported by JSPS KAKENHI Grant Number JPJP24H00017 and JSPS Bilateral Program Number JPJSBP120249910.

E.Pal. acknowledges financial support from the Agencia Estatal de Investigaci\'on of the Ministerio de Ciencia e Innovaci\'on MCIN/AEI/10.13039/501100011033 and the ERDF “A way of making Europe” through project PID2021-125627OB-C32, and from the Centre of Excellence “Severo Ochoa” award to the Instituto de Astrofisica de Canarias.

E.Pak. acknowledges the Europlanet Telescope Network funded by the European Union's Horizon 2020 Research and Innovation Programme (Grant agreement No. 871149).

A. P.-T. acknowledges financial support from the Severo Ochoa grant CEX2021-001131-S funded by MCIN/AEI/ 10.13039/501100011033.

P.P. was partially supported by the project Partnership for Excellence in Superprecise Optics, reg. no. CZ.02.1.01/0.0/0.0/16\_026/0008390.

D.F.R is thankful for the support of the CAPES and FAPERJ/DSC-10 (SEI-260003/001630/2023).

F.S. would like to thank Danilo Zardin for his support in the data reduction and Marco Fiaschi for technical support with the instrumentation.

M.S. and M.Z. were partially supported by a program of the Polish Ministry of Science under the title ‘Regional Excellence Initiative’, project no. RID/SP/0050/2024/1.

M. T. would like to thank Sharjah Astronomical Observatory (SAO-M47) for its full support in using its telescope under the Exoplanet Observation Project.

A.T. acknowledges the support of the Foundation for the Development of Theoretical Physics and Mathematics BASIS (project 24-2-1-6-1). Their observations were performed at telescope Astrosib RC-500-Kislovodsk of shared research facility “Terskol observatory” of Institute of Astronomy of the Russian Academy of Sciences.

AHMJT acknowledged financial support from the European Research Council (ERC) under the European Union's Horizon 2020 research and innovation program (grant agreement No. 803193/BEBOP), from the UKRI Frontiers research grant (EP/Z000327/1/CandY) and from the Science and Technology Facilities Council (STFC; grant No. ST/S00193X/1, ST/W002582/1 and ST/Y001710/1).

K.V. would like to thank the UCLO (University College London Observatory ) team for their constant guidance and help for their contributions to ExoClock.

{\small
\bibliographystyle{aasjournals}
\bibliography{references_literature,references_tab_parameters,references_tab_updated_ephemerides,references}
}

\appendix

\section{SUPPLEMENTARY INFORMATION} \label{sec:sup-info}

Here we append extra information regarding the data sources and results. More specifically, Table \ref{tab:private_observatories} includes a list with the amateur private observatories contributing to this work, and is followed by a description of the ASTEP telescope. Table \ref{tab:parameters} includes a list with the parameters used in the analysis of individual light-curves and the respective references, where the asterisk indicates orbital parameters ($a/R_s$ or $i$) that were adjusted based on TESS data to match the observed durations.



\end{document}